# Automating Induction by Reflection

Johannes Schoisswohl
University of Manchester, UK TU Wien, Austria
johannes.schoisswohl@manchester.ac.uk

Laura Kovács
TU Wien, Austria
laura.kovacs@tuwien.ac.at

Despite recent advances in automating theorem proving in full first-order theories, inductive reasoning still poses a serious challenge to state-of-the-art theorem provers. The reason for that is that in first-order logic induction requires an infinite number of axioms, which is not a feasible input to a computer-aided theorem prover requiring a finite input. Mathematical practice is to specify these infinite sets of axioms as axiom schemes. Unfortunately these schematic definitions cannot be formalized in first-order logic, and therefore not supported as inputs for first-order theorem provers.

In this work we introduce a new method, inspired by the field of axiomatic theories of truth, that allows to express schematic inductive definitions, in the standard syntax of multi-sorted first-order logic. Further we test the practical feasibility of the method with state-of-the-art theorem provers, comparing it to solvers' native techniques for handling induction.

## 1 Introduction

Automated reasoning has advanced tremendously in the last decades, pushing the limits of what computer programs can prove about other computer programs. Recent progress in this area features techniques such as first-order reasoning with term algebras [18], embedding programming control structures in first-order logic [17], the AVATAR architecture [27], combining theory instantiation and unification with abstraction in saturation-based proof search [23], automating proof search for higher-order logic [3, 4], and first-order logic with rank-1 polymorphism [5].

Despite the fact that first-order logic with equality can be handled very well in many practical cases, there is one fundamental mathematical concept most first-order theorem provers lack; namely inductive reasoning. Not only is induction a very interesting theoretical concept, it is also of high practical importance, since it is required to reason about software reliability, safety and security, see e.g. [13, 7, 10, 22, 11]. Such and similar software requirement typically involve properties over natural numbers, recursion, unbounded loop iterations, or recursive data structures like lists, and trees.

Many different approaches towards automating inductive reasoning have been proposed, ranging from cyclic proofs [9, 16], computable approximations of the $\omega$-rule [2], recursion analysis [1, 8, 21], theory exploration [6, 25], inductive strengthening [20, 25], and integrating induction in saturation based theorem proving [12, 8, 24]. What all these approaches have in common is that they tackle the problem of inductive theorem proving by specializing the proof system. In the present paper, we propose a different approach. Instead of changing the proof system, we will change our input problem, replacing the infinite induction scheme by a finite conservative extension. Hence our approach is not tailored to one specific reasoner, but can be used with any automated theorem prover for first-order logic.

In order to achieve such a generic approach, we use ideas from axiomatic theories of truth [15]. As Gödel's incompleteness theorem tells us that there is a formula $Bew(x)$ in Peano Arithmetic **PA** that expresses provability in **PA**, Tarski's undefinability theorem teaches us that there is no formula $T(x)$ that expresses truth in **PA**. Extending the language of **PA**, in order to be able to express truth in **PA** is the core idea of axiomatic theories of truth. The truth theory we introduce in this paper will not be an extension of **PA**, but an extension of an arbitrary theory. This theory will let us express the sentence "for all formulas, the induction scheme is true", within first order logic.

As our experimental evaluation shows our technique does not outperform state of the art built-in methods for inductive reasoning in general, we will see that there are cases where an improvement can be achieved.

The contributions of our work can be summarized as follows:

- We introduce a method for conservatively extending an arbitrary theory with a truth predicate (Section 3);
- We show how our method can be used to replace the induction scheme of **PA**, and other theories involving inductive datatypes (Section 4);
- We provide a new set of benchmarks to the automated theorem proving community (Section 5);
- We conduct experiments on this set of benchmarks with state-of-the-art theorem provers. (Section 5)

## 2 Preliminaries

We assume familiarity with multi-sorted first-order logic and automated theorem proving; for details, we refer to [19, 5].

We consider reasoning on the object and the meta level. Therefore we will use different symbols for logic on all of these levels. We use the symbols $\neg$, $\wedge$, $\vee$, $\rightarrow$, $\leftrightarrow$, and $\approx$ for negation, conjunction, disjunction, implication, equivalence, and equality respectively, and write $Qx : \sigma.\phi$ with $Q \in \{\exists, \forall\}$ for existential and universal quantification over the sort $\sigma$ on the object level. We will drop sort declarations for quantified variables when there is no ambiguity. Further we will write $\bigforall_{x_1,\ldots,x_n} \phi$ for the formula $\forall x_1.\ldots\forall x_n.\phi$, and $\bigforall\phi$ to denote the universal closure of $\phi$.

On the meta-level we will use !, &, ||, $\Longrightarrow$, $\Longleftrightarrow$, and $=$ for negation, conjunction, disjunction, implication, equivalence, and equality and $\forall$, and $\exists$ for quantification. Meta-level logical formulas will only be used where they help to improve readability and precision, and otherwise natural language will be used.

By $\mathbf{Var}_\sigma$, $\mathbf{Term}_\sigma$, and **Form**, we respectively denote the sets of variables of sort $\sigma$, terms of sort $\sigma$ and object-level formulas over a signature $\Sigma$. As $\mathbf{Var}_\sigma$ is a countably infinite set, we assume without loss of generality that it is composed of the variables $\{\mathrm{x}_i^\sigma \mid i \in \mathbb{N}\}$, and leave away the sort superscript $\sigma$, if it is clear from the context.

For a function symbol $f$ from signature $\Sigma$, we write $f :: \sigma_1 \times \ldots \times \sigma_n \rightsquigarrow \sigma \in \Sigma$, to denote that $\mathbf{dom}(f) = \sigma_1 \times \ldots \times \sigma_n$ is the domain of $f$, $\mathbf{codom}(f) = \sigma$ is its codomain, and $\mathbf{arity}(f) = n$ is the arity of $f$. We consider constants being functions of arity 0 and write $c :: \sigma$ for $c :: \rightsquigarrow \sigma$. Further we write $P : \mathsf{Pred}(\sigma_n \times \ldots \times \sigma_n)$ to denote that $P$ is a predicate with domain $\mathbf{dom}(P) = \sigma_1 \times \ldots \times \sigma_n$ and arity $\mathbf{arity}(P) = n$. Further, we write $\mathbf{dom}(s,i)$ to refer to the $i$th component of the domain of $s$.

Given formula/term $\phi$, a variable $x$ and a term $t$, we write $\phi[x \mapsto t]$ to denote the formula/term resulting from replacing all occurences of $x$ by $t$ in $\phi$. Similarly, if $x$ is a variable and $\phi[x]$ is a formula/term, we denote the formula/term resulting from replacing all occurences of $x$ for $t$ by $\phi[t]$.

A formula is open if it contains free variables, and closed otherwise. We consider a theory to be a set of closed formulas. If $\mathscr{T}$ is a theory with signature $\Sigma$, by $\mathbf{Form}^{\mathscr{T}}$ we denote the set of all formulas over $\Sigma$.

The semantics of formulas and terms over a signature $\Sigma$ is defined using multi-sorted first-order interpretations $\mathscr{M}$, consisting of $\langle\langle\Delta_{\sigma_1},\dots,\Delta_{\sigma_n}\rangle,\mathscr{I}\rangle$, where: $\Delta_{\sigma_i}$ is the domain for sort $\sigma \in \mathbf{sorts}_\Sigma$, and $\mathscr{I}$ is an interpretation function that freely interprets variables, function symbols, and predicate symbols, respecting the sorts, and is extended to terms in the standard way. By $\Delta$ we denote $\langle\Delta_{\sigma_1},\dots,\Delta_{\sigma_n}\rangle$. We write $\mathscr{M} \models \phi$ for the structure $\mathscr{M}$ satisfying the formula $\phi$ and say that $\mathscr{M}$ is a model of $\phi$. We write $\mathscr{I} \models \phi$ instead of $\langle\langle\Delta_{\sigma_1},\dots,\Delta_{\sigma_n}\rangle,\mathscr{I}\rangle \models \phi$, whenever the domains are clear from context. By $\mathbf{Interpret}_\Sigma$ we denote the class of all interpretation functions over a signature $\Sigma$.

For defining our approach for reflective reasoning, we need the concept of a conservative extension. A conservative extension of a theory $\mathscr{T}$ is a theory $\mathscr{T}'$, such that $\Sigma_{\mathscr{T}} \subseteq \Sigma_{\mathscr{T}'}$, and $\forall\phi \in \mathbf{Form}^{\mathscr{T}}.(\mathscr{T} \models \phi \iff \mathscr{T}' \models \phi)$

An inductive datatype $\mathscr{D}_\tau$, with respect to some signature $\Sigma$ is a pair $\langle\tau,\mathbf{ctors}_\tau\rangle$, where $\tau$ is a sort and $\mathbf{ctors}_\tau$ is a set of function symbols $F \subset \Sigma$ such that $\forall f \in F.\mathbf{codom}(f) = \tau$. By the first-order structural induction scheme of $\mathscr{D}_\tau$ we denote the set of formulas:

$$\Big\{\big(\bigwedge_{c\in\mathbf{ctors}_\tau} case_c\big) \to \forall x.\phi[x] \mid \phi[x] \in \mathbf{Form}\Big\} \qquad (\mathbf{I}_\tau)$$

where

$$case_c = \bigforall_{x_1,\dots,x_n} \Big(\big(\bigwedge_{i\in recursive_c} \phi[x_i]\big) \to \phi[c(x_1,\dots,x_n)]\Big)$$
$$recursive_c = \{i \mid \mathbf{dom}_\Sigma(c,i) = \tau\}$$

$x$ the induction variable, $\bigwedge_{c\in\mathbf{ctors}_\tau} case_c$ the induction premise, and $\forall x\phi[x]$ the induction conclusion.

An example for such an inductive datatype is the type of lists $\mathscr{D}_{\mathsf{List}} = \langle\mathsf{List},\{\mathsf{nil} :: \mathsf{List},\mathsf{cons} :: \alpha\times \mathsf{List} \rightsquigarrow \mathsf{List}\}\rangle$. The first-order induction scheme for $\mathscr{D}_{\mathsf{List}}$ is therefore

$$\Big\{case_{\mathsf{nil}} \wedge case_{\mathsf{cons}} \to \forall x.\phi[x] \mid \phi[x] \in \mathbf{Form}\Big\} \qquad case_{\mathsf{nil}} = \top \to \phi[\mathsf{nil}]$$
$$case_{\mathsf{cons}} = \forall x : \alpha, xs : \mathsf{List}.\Big(\phi[xs] \to \phi[\mathsf{cons}(x,xs)]\Big)$$

## 3 Reflective extension

Our aim is to finitely axiomatise the induction scheme $\mathbf{I}_\tau$ with respect to some datatypes $\mathscr{D}$ and an arbitrary base theory $\mathscr{T}$. We will hence first construct a conservative extension $\dot{\mathscr{T}}$ of $\mathscr{T}$, which allow us to quantify over first-order formulas of the language of $\mathscr{T}$. In order to achieve this we will take an approach that is inspired by Horsten's theory **TC** [15]. There are however a few crucial differences between our approach and [15], as follows. While **TC** [15] is an extension of Peano Arithmetic **PA**, our work can be used for an arbitrary theory $\mathscr{T}$. Further, while **TC** [15] relies on numbers to encode formulas, our approach uses multi-sorted logic and introduces additional sorts for formulas, terms, and variables, yielding a rather straightforward definition of models and proof of consistency for the extended theory.

The basic idea of our approach is to redefine the syntax and semantics of first-order logic in first-order logic itself. As such, there will be three levels of reasoning. As usual, we have the meta level logic (the informal logical reasoning going on in the paper), and the object level (the formal logical reasoning we reason about on the meta level). In addition we will have a logic embedded in the terms of object level logic formulas. We will refer to this level as the *reflective level*, and refer to formulas/functions/terms/expressions as reflective formulas/functions/terms/expressions.

## 3.1 Signature

In a first step, we extend the signature $\Sigma$ and the set of sorts $\mathbf{sorts}_\Sigma$ of our base theory $\mathscr{T}$ with the vocabulary to be able to talk about variables, terms, and formulas.

**Definition 1** (Reflective signature)**.** *Let $\Sigma$ be an arbitrary signature. We define $\dot\Sigma$ to be the reflective extension of $\Sigma$.*

$$
\begin{aligned}
\mathbf{sorts}_{\dot\Sigma} = {} & \{\mathsf{var}_\sigma \mid \sigma \in \mathbf{sorts}_\Sigma\} \\
& \cup \{\mathsf{term}_\sigma \mid \sigma \in \mathbf{sorts}_\Sigma\} \\
& \cup \{\mathsf{form}, \mathsf{env}\}
\end{aligned}
$$

$$
\begin{aligned}
\dot\Sigma = \Sigma & \cup \{\mathsf{v}_0^\sigma :: \mathsf{var}_\sigma \mid \sigma \in \mathbf{sorts}_\Sigma\} \\
& \cup \{\mathsf{next}_\sigma :: \mathsf{var}_\sigma \rightsquigarrow \mathsf{var}_\sigma \mid \sigma \in \mathbf{sorts}_\Sigma\} \\
& \cup \{\mathsf{inj}_\sigma :: \mathsf{var}_\sigma \rightsquigarrow \mathsf{term}_\sigma \mid \sigma \in \mathbf{sorts}_\Sigma\} \\
& \cup \{\dot f :: \mathsf{term}_{\sigma_1} \times \ldots \times \mathsf{term}_{\sigma_n} \rightsquigarrow \mathsf{term}_\sigma \mid f :: \sigma_1 \times \ldots \times \sigma_n \rightsquigarrow \sigma \in \Sigma\} \\
& \cup \{\dot P :: \mathsf{term}_{\sigma_1} \times \ldots \times \mathsf{term}_{\sigma_n} \rightsquigarrow \mathsf{form} \mid P :: \mathsf{Pred}(\sigma_1 \times \ldots \times \sigma_n) \in \Sigma\} \\
& \cup \{\dot\approx_\sigma :: \mathsf{term}_\sigma \times \mathsf{term}_\sigma \rightsquigarrow \mathsf{form} \mid \sigma \in \mathbf{sorts}_\Sigma\} \\
& \cup \{\dot\bot :: \mathsf{form}, \dot\vee :: \mathsf{form} \times \mathsf{form} \rightsquigarrow \mathsf{form}, \dot\neg :: \mathsf{form} \rightsquigarrow \mathsf{form}\} \\
& \cup \{\dot\forall_\sigma :: \mathsf{var}_\sigma \times \mathsf{form} \rightsquigarrow \mathsf{form} \mid \sigma \in \mathbf{sorts}_\Sigma\} \\
& \cup \{\mathsf{empty} :: \mathsf{env} \mid \sigma \in \mathbf{sorts}_\Sigma\} \\
& \cup \{\mathsf{push}_\sigma :: \mathsf{env} \times \mathsf{var}_\sigma \times \sigma \rightsquigarrow \mathsf{env} \mid \sigma \in \mathbf{sorts}_\Sigma\} \\
& \cup \{\mathsf{eval}^v_\sigma :: \mathsf{env} \times \mathsf{var}_\sigma \rightsquigarrow \sigma \mid \sigma \in \mathbf{sorts}_\Sigma\} \\
& \cup \{\mathsf{eval}_\sigma :: \mathsf{env} \times \mathsf{term}_\sigma \rightsquigarrow \sigma \mid \sigma \in \mathbf{sorts}_\Sigma\} \\
& \cup \{\dot\models :: \mathsf{Pred}(\mathsf{env} \times \mathsf{form})\}
\end{aligned}
$$

*where all newly introduced symbols, and sorts are disjoint from the ones in $\Sigma$, and $\mathbf{sorts}_\Sigma$ respectively.*

As this definition is rather lengthy we will break down the intended semantics of all newly introduced symbols. We can split the definitions into two parts: (i) one formalizing the syntax and (ii) one formalizing the semantics of our reflective first-order logic.

**(i) Reflective syntax.** Our reflective syntax is formalized as follows.

**Variables** The sort $\mathsf{var}_\sigma$ is used to represent the countably infinite set of variables $\mathbf{Var}_\sigma$. The two functions $\mathsf{v}_0^\sigma$, and $\mathsf{next}_\sigma$ that are added to the signature can be thought of as the constructors for this infinite set of variables. This means $\mathsf{v}_0^\sigma$ is intended to be interpreted as the variable $\mathsf{x}_0$, $\mathsf{next}_\sigma(\mathsf{v}_0^\sigma)$ is meant to be interpreted as $\mathsf{x}_1$, and so on. We introduce the following syntactic sugar for variables:

$$\mathsf{v}_{i+1}^\sigma = \mathsf{next}_\sigma(\mathsf{v}_i^\sigma) \qquad\qquad \text{for } i \geq 0$$

**Terms** We use the sort $\mathsf{term}_\sigma$ to represent terms of sort $\mathbf{Term}_\sigma$. On the meta level terms are defined inductively, as follows.

The base case is a variable. Since variables and terms are of different sorts, we need the function $\mathsf{inj}_\sigma$ to turn variables into terms. This function is intended to be interpreted as the identity function.

The step case of the inductive definition is building terms out of function symbols and other terms. Therefore, we need to introduce a reflective function symbol $\dot{f}$ for every function $f$ in the signature. The $\dot{f}$ is intended to be interpreted as the function symbol $f$, while $f$ itself is interpreted as an actual function.

**Formulas** As for terms, formulas **Form** are defined inductively on the meta level.

For atomic formulas we introduce a reflective equality symbol $\dot{\approx}_\sigma$ for each sort $\sigma$ and a reflective version $\dot{P}$ for every predicate symbol $P$. Even though it's not strictly necessary we introduce a nullary reflective connective $\dot{\bot}$ is intended to be interpreted as the formula $\bot$.

Complex formulas are built from atomic formulas and connectives, or quantifiers. Therefore we introduce a functionally complete set of reflective connectives, namely $\dot{\vee}$, and $\dot{\neg}$. As it will help in terms of readability, we will use infix notation for $\dot{\vee}$, and drop the parenthesis for $\dot{\neg}$ if there is no ambiguity.

In order to formalize quantification we introduce a function $\dot{\forall}_\sigma$ for each sort. We will write $\dot{\forall} x{:}\sigma.p$ for the term $\dot{\forall}_\sigma(x,p)$.

**(ii) Reflective semantics.** For axiomatising the meaning of formulas, we will use syntactic representations of the semantic structures needed to define the semantics of first-order logic.

**Environment** In order to define the meaning of a quantifier, we redefine the meaning of a variable within the scope of the quantifier. Therefore we will use a stack of variable interpretations, which we will call an environment. The idea is that a variable $\mathsf{v}_i^\sigma$ is freely interpreted in an empty environment $\mathsf{empty}$, while it is interpreted as $t$ if the tuple $\langle \mathsf{v}_i^\sigma, t\rangle$ was pushed on the stack using $\mathsf{push}_\sigma(e, \mathsf{v}_i^\sigma, t)$. This setting becomes more clear in Sections 3.2-3.3, where we axiomatise the meaning and define a model of the reflective theory.

**Evaluation** To make use of the environment, we need a reflective evaluation function for terms $\mathsf{eval}_\sigma$ and $\mathsf{eval}_\sigma^v$ that corresponds to interpreting terms and variables in some model $\mathscr{I}$ of the reflective theory.

**Satisfaction** Finally, we have our reflective satisfaction relation $\dot{\models}$. We write $e \dot{\models} p$ for $\dot{\models}(e,p)$, which can roughly be interpreted as "the interpretation $\mathscr{I}$ partially defined by $e$ satisfies $p$". Our truth $\mathbf{T}$ predicate in the Tarskian sense is $\mathbf{T}(x) = (\mathsf{empty} \dot{\models} x)$.

## 3.2 Axiomatisation

We now formalize our semantics. We relate reflective with non-reflective function and predicate symbols, by defining the meaning of the reflective satisfaction relation $\dot{\models}$, and the meaning of the reflective evaluation functions $\mathsf{eval}_\sigma$, and $\mathsf{eval}_\sigma^v$. All axioms we list are implicitly universally quantified, and one instance of them will be present for every sort $\sigma, \tau \in \mathbf{sorts}_\Sigma$. Finally, the reflective extension $\dot{\mathscr{T}}$ of our base theory $\mathscr{T}$ is the union of all these axioms and $\mathscr{T}$.

**Reflective variable interpretation.** As already mentioned, the interpretation of variables in an empty environment empty is undefined. In contrast an environment to which a variable $v$, and a value $x$ is pushed, evaluates the variable $v$ to $x$. Hence,

$$\mathsf{eval}^v_\sigma(\mathsf{push}_\sigma(e,v,x),v) = x \qquad (\mathsf{Ax}_{\mathsf{eval}^v_0})$$

$$v \not\approx v' \to \mathsf{eval}^v_\sigma(\mathsf{push}_\sigma(e,v',x),v) = \mathsf{eval}^v_\sigma(e,v) \qquad (\mathsf{Ax}_{\mathsf{eval}^v_1})$$

$$\mathsf{eval}^v_\sigma(\mathsf{push}_\tau(e,w,x),v) = \mathsf{eval}^v_\sigma(e,v) \quad \text{for } \sigma \neq \tau \qquad (\mathsf{Ax}_{\mathsf{eval}^v_2})$$

**Reflective evaluation.** The function symbol $\mathsf{eval}_\sigma$ defines the value of a reflective term $t$, and thereby maps the reflective functions $\dot{f}$ to their non-reflective counter parts $f$. For variables $\mathsf{eval}_\sigma$, the evaluation to $\mathsf{eval}^v_\sigma$ is used.

$$\mathsf{eval}_\sigma(e,\mathsf{inj}_\sigma(v)) = \mathsf{eval}^v_\sigma(e,v) \qquad (\mathsf{Ax}_{\mathsf{eval}_{var}})$$

$$\mathsf{eval}_\sigma(e,\dot{f}(t_1,\dots,t_n)) = f(\mathsf{eval}_{\sigma_1}(e,t_1),\dots,\mathsf{eval}_{\sigma_n}(e,t_n)) \qquad (\mathsf{Ax}_{\mathsf{eval}_f})$$

$$\text{for } f : \sigma_1 \times \dots \times \sigma_n \rightsquigarrow \sigma \in \Sigma$$

**Reflective satisfaction.** The predicate symbol $\dot{\models}\,\sigma$ defines the truth of a formula with respect to some variable interpretation. To this end, the meaning of the reflective connectives and the quantifiers in terms is defined by their respective object-level counterparts, as follows:

$$(e \mathrel{\dot{\models}} x \mathrel{\dot{\approx}_\sigma} y) \leftrightarrow \mathsf{eval}_\sigma(e,x) \approx \mathsf{eval}_\sigma(e,y) \qquad (\mathsf{Ax}_{\dot{\approx}})$$

$$(e \mathrel{\dot{\models}} \dot{P}(t_1,\dots,t_n)) \leftrightarrow P(\mathsf{eval}_{\sigma_1}(e,t_1),\dots,\mathsf{eval}_{\sigma_n}(e,t_n)) \quad \text{for } P : \mathsf{Pred}(\sigma_1 \times \dots \times \sigma_n) \qquad (\mathsf{Ax}_P)$$

$$(e \mathrel{\dot{\models}} \dot{\bot}) \leftrightarrow \bot \qquad (\mathsf{Ax}_{\dot{\bot}})$$

$$(e \mathrel{\dot{\models}} \dot{\neg}\, \phi) \leftrightarrow \neg(e \mathrel{\dot{\models}} \phi) \qquad (\mathsf{Ax}_{\dot{\neg}})$$

$$(e \mathrel{\dot{\models}} \phi \mathrel{\dot{\vee}} \psi) \leftrightarrow (e \mathrel{\dot{\models}} \phi) \vee (e \mathrel{\dot{\models}} \psi) \qquad (\mathsf{Ax}_{\dot{\vee}})$$

$$(e \mathrel{\dot{\models}} \dot{\forall} v{:}\sigma.\phi) \leftrightarrow \forall x : \sigma.(\mathsf{push}_\sigma(e,v,x) \mathrel{\dot{\models}} \phi) \qquad (\mathsf{Ax}_{\dot{\forall}})$$

## 3.3 Consistency and Conservativeness

As we have now specified our theory, we next ensure that (i) $\dot{\mathscr{T}}$ is indeed a conservative extension of $\mathscr{T}$ and (ii) $\dot{\mathscr{T}}$ is consistent. In general, we cannot ensure that $\dot{\mathscr{T}}$ is consistent, since already the base theory $\mathscr{T}$ could have been inconsistent. Hence we will show that $\dot{\mathscr{T}}$ is consistent if $\mathscr{T}$ is consistent.

In order to prove (i) and (ii), that is conservativeness and consistency of $\dot{\mathscr{T}}$, we introduce the notion of a *reflective model* $\dot{\mathscr{M}}$, that is based on a model $\mathscr{M}$ of $\mathscr{T}$. The basic idea is that $\dot{\mathscr{M}}$ interprets every symbol in the base theory $\mathscr{T}$ as it would be interpreted in $\mathscr{M}$, hence every formula in $\mathbf{Form}^{\mathscr{T}}$ is true in $\dot{\mathscr{M}}$ iff it is true in $\mathscr{M}$. Due to soundness and completeness of first-order logic we get that $\dot{\mathscr{T}}$ is indeed a conservative extension of $\mathscr{T}$. Further, due to the fact that for every model of $\mathscr{M}$ of $\mathscr{T}$ we have a model $\dot{\mathscr{M}}$ of $\dot{\mathscr{T}}$, we also have that $\dot{\mathscr{T}}$ is consistent if $\mathscr{T}$ is consistent. In order to ensure this reasoning is correct we need to ensure that $\dot{\mathscr{M}}$ also satisfies the axioms we introduced for reflective theories. This will be done by interpreting the new reflective sort form as the set of first order formulas **Form**, and interpreting the sort $\mathsf{term}_\sigma$ as terms of sort $\mathbf{Term}_\sigma$.

**Definition 2** (Reflective interpretation)**.** *Let $\mathscr{M} = \langle\langle\Delta_{\sigma_1},\dots,\Delta_{\sigma_n}\rangle,\mathscr{I}\rangle$ be a first-order interpretation over the signature $\Sigma$. We define the reflective interpretation $\dot{\mathscr{M}}$ to be*

$$\dot{\mathscr{M}} = \langle\langle \Delta_{\sigma_1}, ..., \Delta_{\sigma_n}, \mathbf{Term}_{\sigma_1}, ...\mathbf{Term}_{\sigma_n}, \mathbf{Form}\rangle, \dot{\mathscr{I}}\rangle$$

$$\dot{\mathscr{I}}(f) : \Delta_{\sigma_1} \times ... \times \Delta_{\sigma_n} \mapsto \Delta_\sigma \qquad \textit{for } f :: \sigma_1 \times ... \times \sigma_n \rightsquigarrow \sigma \in \Sigma$$
$$\dot{\mathscr{I}}(f) = \mathscr{I}(f)$$

$$\dot{\mathscr{I}}(P) : \mathscr{P}(\Delta_{\sigma_1} \times ... \times \Delta_{\sigma_n}) \qquad \textit{for } P :: \mathsf{Pred}(\sigma_1 \times ... \times \sigma_n) \in \Sigma$$
$$\dot{\mathscr{I}}(P) = \mathscr{I}(P)$$

$$\dot{\mathscr{I}}(\mathsf{v}_0^\sigma) : \mathbf{Var}_\sigma \qquad \textit{for } \sigma \in \mathbf{sorts}_\Sigma$$
$$\dot{\mathscr{I}}(\mathsf{v}_0^\sigma) = \mathsf{x}_0$$

$$\dot{\mathscr{I}}(\mathsf{next}_\sigma) : \mathbf{Var}_\sigma \mapsto \mathbf{Var}_\sigma \qquad \textit{for } \sigma \in \mathbf{sorts}_\Sigma$$
$$\dot{\mathscr{I}}(\mathsf{next}_\sigma)(\mathsf{x}_i) = \mathsf{x}_{i+1}$$

$$\dot{\mathscr{I}}(\mathsf{inj}_\sigma) : \mathbf{Var}_\sigma \mapsto \mathbf{Term}_\sigma \qquad \textit{for } \sigma \in \mathbf{sorts}_\Sigma$$
$$\dot{\mathscr{I}}(\mathsf{inj}_\sigma)(x) = x$$

$$\dot{\mathscr{I}}(\dot{f}) : \mathbf{Term}_{\sigma_1} \times ... \times \mathbf{Term}_{\sigma_n} \mapsto \mathbf{Term}_\sigma \qquad \textit{for } f :: \sigma_1 \times ... \times \sigma_n \rightsquigarrow \sigma \in \Sigma$$
$$\dot{\mathscr{I}}(\dot{f})(t_1, ..., t_n) = f(\dot{\mathscr{I}}(t_1), ..., \dot{\mathscr{I}}(t_n))$$

$$\dot{\mathscr{I}}(\dot{P}) : \mathbf{Term}_{\sigma_1} \times ... \times \mathbf{Term}_{\sigma_n} \mapsto \mathbf{Form} \qquad \textit{for } P :: \mathsf{Pred}(\sigma_1 \times ... \times \sigma_n) \in \Sigma$$
$$\dot{\mathscr{I}}(\dot{P})(t_1, ..., t_n) = P(\dot{\mathscr{I}}(t_1), ..., \dot{\mathscr{I}}(t_n))$$

$$\dot{\mathscr{I}}(\dot{\approx}_\sigma) : \mathbf{Term}_\sigma \times \mathbf{Term}_\sigma \mapsto \mathbf{Form} \qquad \textit{for } \sigma \in \mathbf{sorts}_\Sigma$$
$$\dot{\mathscr{I}}(\dot{\approx}_\sigma)(s, t) = \dot{\mathscr{I}}(s) \approx \dot{\mathscr{I}}(t)$$

$$\dot{\mathscr{I}}(\dot{\bot}) : \mathbf{Form}$$
$$\dot{\mathscr{I}}(\dot{\bot}) = \bot$$

$$\dot{\mathscr{I}}(\dot{\vee}) : \mathbf{Form} \times \mathbf{Form} \mapsto \mathbf{Form}$$
$$\dot{\mathscr{I}}(\dot{\vee})(\phi, \psi) = \phi \vee \psi$$

$$\dot{\mathscr{I}}(\dot{\neg}) : \mathbf{Form} \mapsto \mathbf{Form}$$
$$\dot{\mathscr{I}}(\dot{\neg})(\phi) = \neg\phi$$

$$\dot{\mathscr{I}}(\dot{\forall}_\sigma) : \mathbf{Var}_\sigma \times \mathbf{Form} \mapsto \mathbf{Form} \qquad \textit{for } \sigma \in \mathbf{sorts}_\Sigma$$
$$\dot{\mathscr{I}}(\dot{\forall}_\sigma)(\mathsf{x}_i, \phi) = \forall \mathsf{x}_i : \sigma.\phi$$

$$\dot{\mathscr{I}}(\mathsf{empty}) : \mathbf{Interpret}_\Sigma$$
$$\dot{\mathscr{I}}(\mathsf{empty}) = \mathscr{I}$$

$$\dot{\mathscr{I}}(\mathsf{push}_\sigma) : \mathbf{Interpret}_\Sigma \times \mathbf{Var}_\sigma \times \sigma \mapsto \mathbf{Interpret}_\Sigma \qquad \textit{for } \sigma \in \mathbf{sorts}_\Sigma$$
$$\dot{\mathscr{I}}(\mathsf{push}_\sigma)(\mathscr{J}, \mathsf{x}_i, v)(x) = \begin{cases} v & \textit{if } x = \mathsf{x}_i \\ \mathscr{J}(x) & \textit{otherwise} \end{cases}$$

$$\dot{\mathscr{I}}(\mathsf{eval}^v_\sigma) : \mathbf{Interpret}_\Sigma \times \mathbf{Var}_\sigma \mapsto \Delta_\sigma \qquad \textit{for } \sigma \in \mathbf{sorts}_\Sigma$$
$$\dot{\mathscr{I}}(\mathsf{eval}^v_\sigma)(\mathscr{J}, \mathsf{x}_i) = \mathscr{J}(\mathsf{x}_i)$$

$$\dot{\mathscr{I}}(\mathsf{eval}_\sigma) : \mathbf{Interpret}_\Sigma \times \mathbf{Term}_\sigma \mapsto \Delta_\sigma \qquad \textit{for } \sigma \in \mathbf{sorts}_\Sigma$$
$$\dot{\mathscr{I}}(\mathsf{eval}_\sigma)(\mathscr{J}, t) = \mathscr{J}(t)$$

$$\dot{\mathscr{I}}(\dot{\models}) : \mathscr{P}(\mathbf{Interpret}_\Sigma \times \mathbf{Form})$$
$$\dot{\mathscr{I}}(\dot{\models}) = \{\langle \mathscr{J}, \phi \rangle \in \mathbf{Interpret}_\Sigma \times \mathbf{Form} \mid \mathscr{J} \models \phi\}$$

We now need to ensure that our reflective interpretation $\dot{\mathscr{M}}$ is indeed a model of $\dot{\mathscr{T}}$ if $\mathscr{M}$ is a model of $\mathscr{T}$.

**Theorem 1** (Reflective model)**.**

$$\mathscr{M} \models \mathscr{T} \iff \dot{\mathscr{M}} \models \dot{\mathscr{T}}$$

*Proof.* The "$\Longleftarrow$" part of the biconditional is trivial since $\mathscr{T} \subset \dot{\mathscr{T}}$, and $\dot{\mathscr{M}}$ interprets all symbols of the original signature in the same way as $\mathscr{M}$.

For the same reason as before we have that $\mathscr{M} \models \mathscr{T} \implies \dot{\mathscr{M}} \models \mathscr{T}$. Hence we are left to show that $\dot{\mathscr{M}} \vdash \dot{\mathscr{T}} \setminus \mathscr{T}$. This follows from the axioms we introduced in 3.2 in natural language, as well as from the meta level semantics of first-order logic, by also making sure that our meta level and our object level definitions match. □

## 3.4 Truth predicate

We showed that our theory $\dot{\mathscr{T}}$ is indeed a conservative extension of $\mathscr{T}$. Next we prove that $\dot{\mathscr{T}}$ behaves in the way we need it for axiomatising induction. That is, we need to make sure that $\dot{\mathscr{T}}$ has a truth predicate, allowing us to quantify over formulas and thus defining the induction scheme as a single formula.

As in [15], we use a Gödel encoding to state that our theory $\dot{\mathscr{T}}$ has a truth predicate. Usually, a Gödel encoding maps variables, terms, and formulas to numerals. Since our theory $\dot{\mathscr{T}}$ does not necessarily contain number symbols, we need to use a more general notion of a Gödel encoding, namely that it maps variables, terms, and formulas in our base language $\mathbf{Form}^{\mathscr{T}}$ to terms in our extended language $\mathbf{Form}^{\dot{\mathscr{T}}}$. That is, we map formulas $\mathbf{Form}$ to terms of sort $\mathsf{form}$, variables $\mathbf{Var}_\sigma$ to $\mathsf{var}_\sigma$ and $\mathbf{Term}_\sigma$ to $\mathsf{term}_\sigma$. Formally, we define our Gödel encoding as follows:

**Definition 3** (Gödel encoding)**.**

$$\ulcorner \phi \vee \psi \urcorner = \ulcorner \phi \urcorner \dot{\vee} \ulcorner \psi \urcorner \qquad (\mathsf{Gdl}_{\vee})$$

$$\ulcorner \neg \phi \urcorner = \dot{\neg} \ulcorner \phi \urcorner \qquad (\mathsf{Gdl}_{\neg})$$

$$\ulcorner \bot \urcorner = \dot{\bot} \qquad (\mathsf{Gdl}_{\bot})$$

$$\ulcorner \forall \mathsf{x}_i : \sigma . \phi \urcorner = \dot{\forall} \mathsf{v}_i^{\sigma} {:} \sigma . \ulcorner \phi \urcorner \qquad (\mathsf{Gdl}_{\forall})$$

$$\ulcorner \mathsf{x}_n \urcorner = \mathsf{inj}_{\sigma}(\mathsf{v}_n^{\sigma}) \qquad \textit{where } \mathsf{x}_n \in \mathbf{Var}_{\sigma} \qquad (\mathsf{Gdl}_{\mathsf{x}})$$

$$\ulcorner s \approx t \urcorner = \ulcorner s \urcorner \dot{\approx}_{\sigma} \ulcorner t \urcorner \qquad \textit{where } s,t \in \mathbf{Term}_{\sigma} \qquad (\mathsf{Gdl}_{\approx})$$

$$\ulcorner f(t_1,...,t_n) \urcorner = \dot{f}(\ulcorner t_1 \urcorner,...,\ulcorner t_n \urcorner) \qquad (\mathsf{Gdl}_{f})$$

$$\ulcorner P(t_1,...,t_n) \urcorner = \dot{P}(\ulcorner t_1 \urcorner,...,\ulcorner t_n \urcorner) \qquad (\mathsf{Gdl}_{P})$$

With our Gödel encoding at hand, we can now show that $\dot{\mathscr{T}}$ contains a truth predicate $\mathbf{T}[\phi]$ for $\mathscr{T}$, namely the formula $\mathsf{empty} \dot{\models} \ulcorner \phi \urcorner$. To this end, we have the following result.

**Theorem 2** ( Truth Predicate )**.**

$$\forall \phi \in \mathbf{Form}^{\mathscr{T}} . \Big( \dot{\mathscr{T}} \models \phi \leftrightarrow (\mathsf{empty} \dot{\models} \ulcorner \phi \urcorner) \Big)$$

□

In order to prove Theorem 2, we strengthen its statements, such that it holds not only for the empty reflective interpretation empty, but also for every reflective interpretation built from terms empty, and $\lambda e.\mathsf{push}_{\sigma}(e, \mathsf{v}_i^{\sigma}, \mathsf{x}_i^{\sigma})$. This is necessary, as (i) our axiomatisation of $\dot{\forall}_{\sigma}$ defines the meaning of a quantifier in terms of a different reflective environment and (ii) we relate finitary representation of variables (based on $\mathsf{v}_0^{\sigma}$ and $\lambda x.\mathsf{next}_{\sigma}(x)$) to the infinite set of variables used for standard first-order logic syntax. For the detailed proof of Theorem 2, we refer to the extended version of this paper [26].

# 4 Induction by Reflection

We next show how to build a finite theory that entails the first-order induction scheme $\mathbf{I}_{\tau}$. For the sake of simplicity we will first have a look at the rather familiar case of Peano Arithmetic, and present a generalisation of the same approach in the following subsection.

## 4.1 Natural Numbers

In order to finitely axiomatise **PA**, we need a finite fragment of **PA** to start with. The obvious choice is **Q**, which we define as **PA** without the induction formulas. We then build the reflective extension $\dot{\mathbf{Q}}$, with the following two essential properties. First, it has a sort form of formulas, hence we can quantify over this sort. Second, we have a truth predicate $\dot{\mathbf{Q}}$ for **Q**, which means we can represent an arbitrary formula of **Q** in a single term in $\dot{\mathbf{Q}}$.

Now we can define a conservative extension $\ddot{\mathbf{Q}}$ of **PA**. Therefore we will add the following axiom to $\dot{\mathbf{Q}}$; we call this axiom as the *reflective induction axiom*:

$$\forall \phi : \mathsf{form}. \Big( \mathbf{True}[\phi, 0] \wedge \forall n : \mathsf{nat}.(\mathbf{True}[\phi, n] \to \mathbf{True}[\phi, \mathsf{s}n]) \to \forall n : \mathsf{nat}.\mathbf{True}[\phi, n] \Big) \qquad (\dot{\mathbf{I}_{\mathsf{nat}}})$$

where $\mathbf{True}[\phi, n] := (\mathsf{push}_{\mathsf{nat}}(\mathsf{empty}, \mathsf{v}_0^{\mathsf{nat}}, n) \dot{\models} \phi)$. Thus, we define $\ddot{\mathbf{Q}}$ as

$$\ddot{\mathbf{Q}} = \dot{\mathbf{Q}} \cup \{\dot{\mathbf{I}_{\mathsf{nat}}}\}$$

**Theorem 3.** $\ddot{\mathbf{Q}}$ *is a conservative extension of* **PA**

*Proof.* We will need the following auxiliary formula:

$$\dot{\mathscr{T}} \models (\mathsf{push}_\sigma(e, \ulcorner \mathsf{x}_i \urcorner, t) \dot{\models} \ulcorner \phi[\mathsf{x}_i] \urcorner) \leftrightarrow (e \dot{\models} \ulcorner \phi[t] \urcorner) \tag{1}$$

This formula holds by induction over $\phi$. Proving that $\ddot{\mathbf{Q}}$ is a conservative extension of **PA** reduces showing that every formula in $\mathbf{Form}^{\mathbf{PA}}$ is provable in $\ddot{\mathbf{Q}}$ iff it is provable in **PA**. We next prove both directions of this property.

**(1)** $\forall \phi \in \mathbf{Form}^{\mathbf{PA}}.(\mathbf{PA} \models \phi \implies \ddot{\mathbf{Q}} \models \phi)$ To this end, we show that all axioms of **PA** are derivable in $\ddot{\mathbf{Q}}$. Since **Q** is a subset of both **PA** and $\ddot{\mathbf{Q}}$, we only need to deal with the induction axioms. Let $\phi[0] \wedge \forall n.(\phi[n] \rightarrow \phi[n+1]) \rightarrow \forall n.\phi[n]$ be an arbitrary instance of the first-order mathematical induction scheme $\mathbf{I}_{\mathsf{nat}}$. Let us instantiate the reflective induction axiom ($\dot{\mathbf{I}_{\mathsf{nat}}}$) with $\ulcorner \phi[\mathsf{x}_0] \urcorner$. We obtain

$$\begin{aligned} \ddot{\mathbf{Q}} \vdash & \mathbf{True}[\ulcorner \phi[\mathsf{x}_0] \urcorner, 0] \wedge \\ & \quad \forall n : \mathsf{nat}.(\mathbf{True}[\ulcorner \phi[\mathsf{x}_0] \urcorner, n] \rightarrow \mathbf{True}[\ulcorner \phi[\mathsf{x}_0] \urcorner, \mathsf{s}n]) \\ & \quad \rightarrow \forall n : \mathsf{nat}.\mathbf{True}[\ulcorner \phi[\mathsf{x}_0] \urcorner, n] \end{aligned}$$

which expands to

$$\begin{aligned} \ddot{\mathbf{Q}} \vdash & (\mathsf{push}_{\mathsf{nat}}(\mathsf{empty}, \mathsf{v}_0^{\mathsf{nat}}, 0) \dot{\models} \ulcorner \phi[\mathsf{x}_0] \urcorner) \wedge \\ & \quad \forall n : \mathsf{nat}.((\mathsf{push}_{\mathsf{nat}}(\mathsf{empty}, \mathsf{v}_0^{\mathsf{nat}}, n) \dot{\models} \ulcorner \phi[\mathsf{x}_0] \urcorner) \rightarrow (\mathsf{push}_{\mathsf{nat}}(\mathsf{empty}, \mathsf{v}_0^{\mathsf{nat}}, \mathsf{s}n) \dot{\models} \ulcorner \phi[\mathsf{x}_0] \urcorner)) \\ & \rightarrow \forall n : \mathsf{nat}.(\mathsf{push}_{\mathsf{nat}}(\mathsf{empty}, \mathsf{v}_0^{\mathsf{nat}}, n) \dot{\models} \ulcorner \phi[\mathsf{x}_0] \urcorner) \end{aligned}$$

By formula (1), we can derive

$$\begin{aligned} \ddot{\mathbf{Q}} \vdash & (\mathsf{empty} \dot{\models} \ulcorner \phi[0] \urcorner) \wedge \\ & \quad \forall n : \mathsf{nat}.((\mathsf{empty} \dot{\models} \ulcorner \phi[n] \urcorner) \rightarrow (\mathsf{empty} \dot{\models} \ulcorner \phi[\mathsf{s}n] \urcorner)) \\ & \rightarrow \forall n : \mathsf{nat}.(\mathsf{empty} \dot{\models} \ulcorner \phi[n] \urcorner) \end{aligned}$$

Applying Theorem 2, the fact that $\lambda x.(\mathsf{empty} \dot{\models} x)$ is our truth predicate, we finally get

$$\ddot{\mathbf{Q}} \vdash \phi[0] \wedge \forall n : \mathsf{nat}.(\phi[n] \rightarrow \phi[\mathsf{s}n]) \rightarrow \forall n : \mathsf{nat}.\phi[n]$$

**(2)** $\forall \phi.(\ddot{\mathbf{Q}} \models \phi \implies \mathbf{PA} \models \phi)$ We prove by contraposition. Suppose we have some formula $\phi$ such that $\mathbf{PA} \not\models \phi$. Hence there is a counter-model $\mathscr{M}$, such that $\mathscr{M} \models \mathbf{PA}$ but $\mathscr{M} \not\models \phi$. Since $\mathbf{Q} \subset \mathbf{PA}$, it holds that $\mathscr{M} \models \mathbf{Q}$. Thanks to Section 3.3 we can extend the model $\mathscr{M}$ to the reflective model $\dot{\mathscr{M}}$ such that $\dot{\mathscr{M}} \models \dot{\mathbf{Q}}$, and that $\dot{\mathscr{M}} \not\models \phi$. We are thus left with establish that $\dot{\mathscr{M}}$ is a model of $\ddot{\mathbf{Q}}$.

In $\dot{\mathcal{M}}$ the sort form is interpreted as the actual set of formulas **Form**. Therefore let $\phi[\mathsf{x}_0]$ be an arbitrary of these formulas. Since $\mathcal{M} \models \mathbf{PA}$, we have that $\dot{\mathcal{M}} \models \mathbf{PA}$, which implies that

$$\dot{\mathcal{M}} \models \phi[0] \land \forall n : \mathsf{nat}.(\phi[n] \rightarrow \phi[\mathsf{s}n]) \rightarrow \forall n : \mathsf{nat}.\phi[n]$$

By Theorem 2, we get

$$\begin{aligned} \dot{\mathcal{M}} \models &(\mathsf{empty} \mathrel{\dot{\models}} \ulcorner\phi[0]\urcorner) \land \\ &\quad \forall n : \mathsf{nat}.((\mathsf{empty} \mathrel{\dot{\models}} \ulcorner\phi[n]\urcorner) \rightarrow (\mathsf{empty} \mathrel{\dot{\models}} \ulcorner\phi[\mathsf{s}n]\urcorner)) \\ \rightarrow &\forall n : \mathsf{nat}.(\mathsf{empty} \mathrel{\dot{\models}} \ulcorner\phi[n]\urcorner) \end{aligned}$$

which, using formula 1, can be rewritten to

$$\begin{aligned} \dot{\mathcal{M}} \models &\mathbf{True}[\phi[x_0], 0] \land \\ &\quad \forall n : \mathsf{nat}.(\mathbf{True}[\phi[x_0], n] \rightarrow \mathbf{True}[\phi[x_0], \mathsf{s}n]) \\ \rightarrow &\forall n : \mathsf{nat}.\mathbf{True}[\phi[x_0], n] \end{aligned}$$

Since $\dot{\mathcal{M}}$ interprets form formulas exactly as the set **Form** and $\phi[x_0]$ is an arbitrary formula, we conclude that the reflective induction axiom $\dot{\mathbf{I}}_\tau$ holds for $\dot{\mathcal{M}}$. Therefore, $\dot{\mathcal{M}}$ models $\ddot{\mathbf{Q}}$ but not $\phi$, which concludes our proof. □

### 4.2 Arbitrary datatypes

The result of Section 4.1 can be lifted to arbitrary datatypes. Therefore, we translate the meta-level definition of the induction scheme $\mathbf{I}_\tau$ for datatypes $\mathcal{D}_\tau$ to an equivalent reflective version. That is, for a theory $\mathcal{T}$, we build $\mathcal{T}'$ by adding the axiom $\dot{\mathbf{I}}_\tau$ to $\dot{\mathcal{T}}$ for every datatype $\mathcal{D}_\tau$ in the theory, as follows:

$$\forall \phi : \mathsf{form}.\Big( \bigwedge_{c \in \mathbf{ctors}} case_{\phi,c} \rightarrow \forall x : \tau.\mathbf{True}[\phi, x]\Big) \qquad (\dot{\mathbf{I}}_\tau)$$

where

$$\begin{aligned} case_{\phi,c} &:= \dot{\forall}_{x_1,\ldots,x_n} \Big( \bigwedge_{i \in recursive_c} \mathbf{True}[\phi, \mathsf{x}_i] \rightarrow \mathbf{True}[\phi, c(x_1, \ldots, x_n)]\Big) \\ recursive_c &:= \{i \mid \mathbf{dom}_\Sigma(c, i) = \tau\} \\ \mathbf{True}[\phi, n] &:= (\mathsf{push}_\tau(\mathsf{empty}, \mathsf{v}_0^\tau, n) \mathrel{\dot{\models}} \phi) \end{aligned}$$

In the case of extending $\mathbf{Q}$ to a conservative extension of $\mathbf{PA}$, the axioms of constructor disjointness $\mathsf{Disj}_{\mathsf{nat}}$, and injectivity $\mathsf{Inj}_{\mathsf{nat}}$ were already present in $\mathbf{Q}$. Thus, for an arbitrary inductive theory $\mathcal{T}$ with inductive datatypes $\mathcal{D}_{\mathcal{T}}$, we define the reflective inductive extension $\ddot{\mathcal{T}}$ as follows:

$$\ddot{\mathcal{T}} = \mathcal{T} \cup \{(\dot{\mathbf{I}}_\tau), \mathsf{Disj}_\tau, \mathsf{Inj}_\tau \mid \mathcal{D}_\tau \in \mathcal{D}_{\mathcal{T}}\}$$

# 5 Experiments

In order to evaluate the practical viability of the techniques introduced Sections 3-4, we performed two set of experiments, denoted as **Refl** and **Ind** and described next.

**Setup** Note that our work introduces many new function symbols, and axioms which might blow up the proof search space, even if induction is not involved at all. Therefore, in our first experiment **Refl** we wanted to evaluate the feasibility of reasoning in the reflective extension of a theory.

**Refl** itself consists of two groups of benchmarks $\mathbf{Refl_0}$, and $\mathbf{Refl_1}$. **Refl** is the simplest one. For every theory $\mathscr{T}$ in some set of base theories, and every axiom $\alpha \in \mathscr{T}$ we try to proof the validity of $\dot{\mathscr{T}} \vdash (\mathsf{empty} \dot{\models} \ulcorner\alpha\urcorner)$. Since we established that $\lambda x.(\mathsf{empty} \dot{\models} x)$ is the truth predicate of $\mathscr{T}$, and the fact that $\alpha$ is an axiom, we know that these consequence assertions indeed hold. $\mathbf{Refl_1}$ involves reasoning in the reflective extension $\dot{\mathscr{T}}$ of some theory as well. But in this case not the reflective version of the axioms, but the reflective versions of some simple consequence of $\mathscr{T}$ are to be proven.

The benchmarks in the second experiment **Ind** are a set of crafted properties that require inductive reasoning. Every problem $\mathscr{T} \models \phi$ in this set of benchmarks is addressed in two ways. Firstly, proving it directly for the solvers that support induction natively, and secondly, translating the problem to $\ddot{\mathscr{T}} \models \phi$. For a description of the exact axioms and conjectures used in each of our benchmarks we refer to the extended version of this paper [26].

All benchmarks, as well as a program for generating reflective, and reflective inductive extensions of theories, and Gödel encodings for conjectures can be found at GITHUB[1]. As the different solvers we used for evaluation support different input formats, our tool supports serializing problems into these various formats.

We used two (non-disjoint) sets of solvers. Firstly, solvers that support induction natively, and secondly various general-purpose theorem provers that are able to deal with multi-sorted quantified first-order logic, hence induction using the reflective extension.

The solvers considered where the SMT-solvers CVC4 and Z3, the superposition-based first-order theorem prover VAMPIRE, the higher-order theorem prover ZIPPERPOSITION that uses a combination of superposition and term rewriting, and the inductive theorem prover ZENO, that is designed to proof inductive properties of a HASKELL-like programming language. Since VAMPIRE in many cases uses incomplete strategy, per default it was run with a complete strategy forced as well. This configuration if referred to as VAMPIRECOMPLETE. ZIPPERPOSITION supports replacing equalities by dedicated rewrite rules, which comes at the cost of the theoretical loss of some provable problems, but yields a significant gain of performance in practice. ZIPPERPOSITION with these rewrite rules enabled will be referred to as ZIPREWRITE. CVC4 allows for theory exploration which was shown to be helpful for inductive reasoning in [12]. CVC4 with this heuristic enabled is referred to as CVC4GEN.

We ran each solver with a timeout of 10 seconds per problem.

**Results** In the first part of Table 1 we can see the results of solvers proving reflective versions of axioms. What is striking is that the SMT solvers CVC4, and Z3, can solver all benchmarks of this category, while the problem seems to be harder for the saturation based theorem provers. Further ZIPREWRITE does pretty well in this class of benchmarks as well. A potential reason for this difference in performance between the ordinary saturation approach and ZIPREWRITE might have to do with the following: For ZIPREWRITE equalities for function definitions of the reflective extensions are translated to rewrite rules that are oriented in way that they would intuitively be oriented by a human, this means that for example the axiom $(\mathsf{Ax}_{\mathsf{eval}_f})$ can be evaluated as one would intuitively do. In contrast VAMPIRE, using superposition with the Knuth-Bendix simplification ordering will orient this equality in the wrong way, which means that it won't be able to evaluate it in the intuitive way, which might be the reason for the difference in performance.

[1] `https://github.com/joe-hauns/msc-automating-induction-via-reflection`

| benchmark | Cvc4 | Cvc4Gen | Z3 | Vampire | VampireComplete | ZipperPosition | ZipRewrite |
|---|---|---|---|---|---|---|---|
| **Refl$_0$** | | | | | | | |
| `N+Leq+Add+Mul-ax0` | ✓ | ✓ | ✓ | ✓ | ✓ | ✓ | ✓ |
| `N+Leq+Add+Mul-ax1` | ✓ | ✓ | ✓ | – | – | – | ✓ |
| `N+Leq+Add+Mul-ax2` | ✓ | ✓ | ✓ | ✓ | ✓ | – | ✓ |
| `N+Leq+Add+Mul-ax3` | ✓ | ✓ | ✓ | – | – | – | ✓ |
| `N+Leq+Add+Mul-ax4` | ✓ | ✓ | ✓ | ✓ | ✓ | – | ✓ |
| `N+Leq+Add+Mul-ax5` | ✓ | ✓ | ✓ | – | – | – | ✓ |
| `N+L+Pref+App-ax0` | ✓ | ✓ | ✓ | ✓ | ✓ | ✓ | ✓ |
| `N+L+Pref+App-ax1` | ✓ | ✓ | ✓ | ✓ | ✓ | – | ✓ |
| `N+L+Pref+App-ax2` | ✓ | ✓ | ✓ | – | – | – | – |
| `N+L+Pref+App-ax3` | ✓ | ✓ | ✓ | ✓ | ✓ | – | ✓ |
| `N+L+Pref+App-ax4` | ✓ | ✓ | ✓ | – | – | – | ✓ |
| **Refl$_1$** | | | | | | | |
| `eqRefl` | ✓ | ✓ | ✓ | ✓ | ✓ | ✓ | ✓ |
| `eqTrans` | ✓ | ✓ | ✓ | – | – | – | ✓ |
| `excludedMiddle-0` | ✓ | ✓ | ✓ | ✓ | ✓ | ✓ | ✓ |
| `excludedMiddle-1` | ✓ | ✓ | ✓ | ✓ | ✓ | ✓ | ✓ |
| `universalInstance` | – | – | ✓ | ✓ | ✓ | ✓ | ✓ |
| `contraposition-0` | ✓ | ✓ | ✓ | ✓ | ✓ | ✓ | ✓ |
| `contraposition-1` | ✓ | ✓ | ✓ | – | – | – | ✓ |
| `currying-0` | ✓ | ✓ | ✓ | ✓ | ✓ | – | ✓ |
| `currying-1` | ✓ | ✓ | ✓ | – | – | – | ✓ |
| `addGround-0` | ✓ | ✓ | ✓ | ✓ | ✓ | ✓ | ✓ |
| `addGround-1` | ✓ | ✓ | – | – | – | ✓ | ✓ |
| `addExists` | – | – | – | – | – | – | ✓ |
| `existsZeroAdd` | – | – | – | – | – | – | – |
| `mulGround` | ✓ | ✓ | ✓ | – | – | – | ✓ |
| `mulExists` | – | – | – | – | – | – | ✓ |
| `existsZeroMul` | – | – | – | – | – | – | – |
| `appendGround-0` | ✓ | ✓ | ✓ | ✓ | ✓ | ✓ | ✓ |
| `appendGround-1` | ✓ | ✓ | ✓ | – | – | ✓ | ✓ |
| `appendExists` | – | – | – | – | – | – | ✓ |
| `existsNil` | – | – | ✓ | ✓ | ✓ | – | ✓ |

Table 1: Results of the experiment **Refl**.

| benchmark | Cvc4 | Cvc4Gen | Vampire | VampireComplete | ZipPerposition | ZipRewrite | Zeno | $\ddot{\text{Cvc4}}$ | $\ddot{\text{Cvc4Gen}}$ | $\ddot{\text{Z3}}$ | $\ddot{\text{Vampire}}$ | $\ddot{\text{VampireComplete}}$ | $\ddot{\text{ZipPerposition}}$ | $\ddot{\text{ZipRewrite}}$ |
|---|---|---|---|---|---|---|---|---|---|---|---|---|---|---|
| `addCommut` | ✓ | ✓ | ✓ | ✓ | ✓ | ✓ | ✓ | – | – | – | – | – | – | – |
| `mulCommut` | – | – | – | – | – | – | – | – | – | – | – | – | – | – |
| `addAssoc` | ✓ | ✓ | ✓ | ✓ | ✓ | ✓ | ✓ | – | – | – | – | – | – | – |
| `mulAssoc` | – | – | – | – | – | – | – | – | – | – | – | – | – | – |
| `addNeutral` | ✓ | ✓ | ✓ | ✓ | ✓ | ✓ | ✓ | – | – | – | – | – | – | – |
| `addNeutral-0` | ✓ | ✓ | ✓ | ✓ | ✓ | ✓ | ✓ | – | – | – | – | – | – | – |
| `addNeutral-1` | ✓ | ✓ | ✓ | ✓ | ✓ | ✓ | ✓ | – | – | – | – | – | – | – |
| `mulZero` | ✓ | ✓ | ✓ | ✓ | ✓ | ✓ | ✓ | – | – | – | – | – | – | ✓ |
| `distr-0` | – | – | – | – | – | – | – | – | – | – | – | – | – | – |
| `distr-1` | – | – | – | – | – | ✓ | – | – | – | – | – | – | – | – |
| `leqTrans` | – | – | – | – | – | – |  | – | – | – | – | – | – | – |
| `zeroMin` | ✓ | ✓ | ✓ | ✓ | ✓ | ✓ |  | – | – | ✓ | ✓ | – | – | ✓ |
| `addMonoton-0` | – | – | – | – | – | – |  | – | – | – | – | – | – | – |
| `addMonoton-1` | – | – | – | – | – | – |  | – | – | – | – | – | – | – |
| `addCommutId` | – | ✓ | ✓ | ✓ | ✓ | ✓ | ✓ | – | – | – | – | – | – | – |
| `appendAssoc` | ✓ | ✓ | ✓ | ✓ | ✓ | ✓ | ✓ | – | – | – | – | – | – | – |
| `appendMonoton` | ✓ | ✓ | ✓ | ✓ | ✓ | ✓ | ✓ | – | – | – | – | – | – | – |
| `allEqRefl` | ✓ | ✓ | ✓ | ✓ | ✓ | ✓ | – | – | – | ✓ | – | – | – | – |
| `allEqDefsEquality` | ✓ | ✓ | – | – | ✓ | ✓ | – | – | – | – | – | – | – | – |
| `revSelfInvers` | – | – | – | – | ✓ | – | – | – | – | – | – | – | – | – |
| `revAppend-0` | – | – | – | – | – | ✓ | – | – | – | – | – | – | – | – |
| `revAppend-1` | – | – | – | – | – | ✓ | – | – | – | – | – | – | – | – |
| `revsEqual` | – | – | – | – | – | – | – | – | – | – | – | – | – | – |

Table 2: Lists the results of running solvers on the benchmark set **Ind**. For every solver Slvr that supports full first-order logic with equality as input, there is a solver $\ddot{\text{Slvr}}$ using the reflective inductive theory as an input instead of using the solvers native handling of induction. The greyed out cells mean that the problem cannot be translated to the solvers input format.

The second part of the table shows that the performance of the SMT-solvers drops as soon as more complex reasoning is involved. Especially the problems with conjectures involving existential quantification[2] are hardly solved by the SMT solvers. This is not surprising since SMT solvers target at solving quantifier-free fragments of first-order logic.

Table 2 lists the results of the final experiment **Ind**. As the first experiments have shown reasoning in the reflective theories is hard even for very simple conjectures, it is not surprising that it is even harder for problems that require inductive reasoning to solve. Nevertheless there are some problems that can be solved using the reflective inductive extension instead of built-in induction heuristics. The most striking result is that Z3 is able to solve benchmarks that involve induction, even though it is a SMT-solver without any support for inductive reasoning.

[2]These problem ids contain the substring "exists" in their id.

## 6 Conclusion

It is mathematical practice to define infinite sets of axioms as schemes of formulas. Alas these schemes of axioms are not part of standard input syntax of today's theorem proves. In order to circumvent this shortcoming, we developed a method to express these schematic definitions in the language of first-order logic by means of a conservative extension, which we called the reflective extension of a theory. We showed that this reflective extension is indeed a conservative extension of the base theory. It contains a truth predicate which allows us to quantify over formulas within the language of first-order logic.

We replaced the first-order induction scheme of **PA** by the axioms needed for the reflective extension and a single additional axiom, called the reflective induction axiom. We proved that the resulting theory is indeed a conservative extension of **PA**. Further, we demonstrated how to replace the induction scheme of a theory with arbitrary inductive datatypes. This kind of conservative extension is what we called the reflective inductive extension.

Our experiments show that reasoning in the reflective extension of a theory is hard for modern theorem provers, even for very simple problems. Despite the poor performance in general, we have a positive result serving as a proof of concept of our method, namely that the SMT-solver Z3, which does not support induction natively was able to solve problems that require inductive reasoning.

Investigating our encoding in relation with the proof systems supported by the Dedukti framework [14] is an interesting line for further work. Further we are interested to explore which different proof search heuristics can be used to make our technique feasible for practical applications.

**Acknowledgements** This work has been supported by the ERC consolidator grant 2020 ARTIST 101002685, the ERC starting grant 2014 SYMCAR 639270, the EPSRC grant EP/P03408X/1, and the ERC proof of concept grant 2018 SYMELS 842066.

## References

[1] Raymond Aubin (1979): *Mechanizing Structural Induction Part II: Strategies*. Theor. Comput. Sci. 9, pp. 347–362, doi:10.1016/0304-3975(79)90035-5.

[2] Siani Baker, Andrew Ireland & Alan Smaill (1992): *On the Use of the Constructive Omega-Rule within Automated Deduction*. In: LPAR'92, Lecture Notes in Computer Science 624, Springer, pp. 214–225, doi:10.1007/BFb0013063.

[3] Alexander Bentkamp, Jasmin Christian Blanchette, Simon Cruanes & Uwe Waldmann (2018): *Superposition for Lambda-Free Higher-Order Logic*. In: IJCAR, Lecture Notes in Computer Science 10900, Springer, pp. 28–46, doi:10.1007/978-3-319-94205-6_3.

[4] Ahmed Bhayat & Giles Reger (2020): *A Combinator-Based Superposition Calculus for Higher-Order Logic*. In: IJCAR, Lecture Notes in Computer Science 12166, Springer, pp. 278–296, doi:10.1007/978-3-030-51074-9_16.

[5] Ahmed Bhayat & Giles Reger (2020): *A Polymorphic Vampire - (Short Paper)*. In: IJCAR, Lecture Notes in Computer Science 12167, Springer, pp. 361–368, doi:10.1007/978-3-030-51054-1_21.

[6] Koen Claessen, Moa Johansson, Dan Rosén & Nicholas Smallbone (2012): *HipSpec: Automating Inductive Proofs of Program Properties*. In: ATx'12/WInG'12, EPiC Series in Computing 17, EasyChair, pp. 16–25. Available at `https://easychair.org/publications/paper/Kb7`.

[7] Véronique Cortier, Niklas Grimm, Joseph Lallemand & Matteo Maffei (2018): *Equivalence Properties by Typing in Cryptographic Branching Protocols*. In: POST, Lecture Notes in Computer Science 10804, Springer, pp. 160–187, doi:10.1007/978-3-319-89722-6_7.

[8] Simon Cruanes (2017): *Superposition with Structural Induction*. In: *FroCoS*, *Lecture Notes in Computer Science* 10483, Springer, pp. 172–188, doi:10.1007/978-3-319-66167-4_10.

[9] Mnacho Echenim & Nicolas Peltier (2020): *Combining Induction and Saturation-Based Theorem Proving*. *J. Autom. Reason.* 64(2), pp. 253–294, doi:10.1007/s10817-019-09519-x.

[10] Yotam M. Y. Feldman, James R. Wilcox, Sharon Shoham & Mooly Sagiv (2019): *Inferring Inductive Invariants from Phase Structures*. In: *CAV*, *Lecture Notes in Computer Science* 11562, Springer, pp. 405–425, doi:10.1007/978-3-030-25543-5_23.

[11] Pamina Georgiou, Bernhard Gleiss & Laura Kovács (2020): *Trace Logic for Inductive Loop Reasoning*. *CoRR* abs/2008.01387, doi:10.34727/2020/isbn.978-3-85448-042-6_33.

[12] Márton Hajdú, Petra Hozzová, Laura Kovács, Johannes Schoisswohl & Andrei Voronkov (2020): *Induction with Generalization in Superposition Reasoning*. In: *CICM*, *Lecture Notes in Computer Science* 12236, Springer, pp. 123–137, doi:10.1007/978-3-030-53518-6_8.

[13] Krystof Hoder, Nikolaj Bjørner & Leonardo Mendonça de Moura (2011): *μZ- An Efficient Engine for Fixed Points with Constraints*. In: *CAV*, *Lecture Notes in Computer Science* 6806, Springer, pp. 457–462, doi:10.1007/978-3-642-22110-1_36.

[14] Gabriel Hondet & Frédéric Blanqui (2020): *The New Rewriting Engine of Dedukti (System Description)*. In Zena M. Ariola, editor: *FSCD*, *LIPIcs* 167, Schloss Dagstuhl - Leibniz-Zentrum für Informatik, pp. 35:1–35:16, doi:10.4230/LIPIcs.FSCD.2020.35.

[15] Leon Horsten (2011): *The Tarskian Turn: Deflationism and Axiomatic Truth*. Mit Press, MIT Press, doi:10.7551/mitpress/9780262015868.001.0001.

[16] Abdelkader Kersani & Nicolas Peltier (2013): *Combining Superposition and Induction: A Practical Realization*. In: *FroCoS*, *Lecture Notes in Computer Science* 8152, Springer, pp. 7–22, doi:10.1007/978-3-642-40885-4_2.

[17] Evgenii Kotelnikov, Laura Kovács, Giles Reger & Andrei Voronkov (2016): *The vampire and the FOOL*. In: *CPP*, ACM, pp. 37–48, doi:10.1145/2854065.2854071.

[18] Laura Kovács, Simon Robillard & Andrei Voronkov (2017): *Coming to terms with quantified reasoning*. In: *POPL*, ACM, pp. 260–270, doi:10.1145/3009837.3009887.

[19] Laura Kovács & Andrei Voronkov (2013): *First-Order Theorem Proving and Vampire*. In: *CAV*, *Lecture Notes in Computer Science* 8044, Springer, pp. 1–35, doi:10.1007/978-3-642-39799-8_1.

[20] K. Rustan M. Leino (2012): *Automating Induction with an SMT Solver*. In: *VMCAI*, *Lecture Notes in Computer Science* 7148, Springer, pp. 315–331, doi:10.1007/978-3-642-27940-9_21.

[21] J. Strother Moore (2019): *Milestones from the Pure Lisp theorem prover to ACL2*. *Formal Aspects Comput.* 31(6), pp. 699–732, doi:10.1007/s00165-019-00490-3.

[22] Lauren Pick, Grigory Fedyukovich & Aarti Gupta (2020): *Automating Modular Verification of Secure Information Flow*. In: *FMCAD*, IEEE, pp. 158–168, doi:10.34727/2020/isbn.978-3-85448-042-6_23.

[23] Giles Reger, Martin Suda & Andrei Voronkov (2018): *Unification with Abstraction and Theory Instantiation in Saturation-Based Reasoning*. In: *TACAS*, *Lecture Notes in Computer Science* 10805, Springer, pp. 3–22, doi:10.1007/978-3-319-89960-2_1.

[24] Giles Reger & Andrei Voronkov (2019): *Induction in Saturation-Based Proof Search*. In: *CADE*, *Lecture Notes in Computer Science* 11716, Springer, pp. 477–494, doi:10.1007/978-3-030-29436-6_28.

[25] Andrew Reynolds & Viktor Kuncak (2015): *Induction for SMT Solvers*. In: *VMCAI*, *Lecture Notes in Computer Science* 8931, Springer, pp. 80–98, doi:10.1007/978-3-662-46081-8_5.

[26] Johannes Schoisswohl & Laura Kovacs (2021): *Automating Induction by Reflection*. Available at `https://arxiv.org/abs/2106.05066`.

[27] Andrei Voronkov (2014): *AVATAR: The Architecture for First-Order Theorem Provers*. In: *CAV*, *Lecture Notes in Computer Science* 8559, Springer, pp. 696–710, doi:10.1007/978-3-319-08867-9_46.